\documentclass[usenatbib,twocolumn,useAMS]{mn2e}
\usepackage{graphicx}
\usepackage{latexsym}
\usepackage{dcolumn}% Align table columns on decimal point
\usepackage{bm}% bold math
\input{epsf}
\newcommand{\be}{\begin{equation}}
\newcommand{\ee}{\end{equation}}
\newcommand{\ba}{\begin{eqnarray}}
\newcommand{\ea}{\end{eqnarray}}

\newcommand{\bi}{\begin{itemize}}
\newcommand{\ei}{\end{itemize}}
\newcommand{\bfi}{\begin{figure}
\epsfxsize=9cm
\epsffile}
\newcommand{\efi}{\end{figure}}

\newcommand{\mnras}{MNRAS}
\newcommand{\apj}{ApJ}

\newcommand{\prd}{PRD}

\title[Searching for a cosmological preferred direction]{Searching for
a preferred direction with Union2.1 data}
\author[Xiaofeng Yang, F. Y. Wang \& Zhe Chu]
{Xiaofeng Yang$^{1,2,3}$\thanks{E-mail:xfyang@nju.edu.cn},
F. Y. Wang$^{1,2}$\thanks{E-mail:fayinwang@nju.edu.cn}, Zhe Chu$^{4}$\\
$1$ School of Astronomy and Space Science, Nanjing University, Nanjing, 210093, China\\
$2$ Key Laboratory of Modern Astronomy and Astrophysics (Nanjing University), Ministry of Education, Nanjing 210093, China\\
$3$ State Key Laboratory of Frontiers in Theoretical Physics,
Institute of Theoretical Physics, Chinese Academy of Sciences, Beijing, 100190, China\\
$4$ Key Laboratory for Research in Galaxies and Cosmology, Shanghai
Astronomical Observatory, Chinese Academy of Sciences, \\Nandan Road
80, Shanghai, 200030, China}

\begin{document}
\maketitle

\begin{abstract}
A cosmological preferred direction was reported from the
type Ia supernovae (SNe Ia) data in recent years. We use the
Union2.1 data to give a simple classification of such studies for
the first time. Because the maximum anisotropic direction is
independent of isotropic dark energy models, we adopt two
cosmological models ($\Lambda$CDM, $w$CDM) for the hemisphere
comparison analysis and $\Lambda$CDM model for dipole fit approach.
In hemisphere comparison method, the matter density and the equation
of state of dark energy are adopted as the diagnostic qualities in
the $\Lambda$CDM model and $w$CDM model, respectively. In dipole fit
approach, we fit the fluctuation of distance modulus. We find that
there is a null signal for the hemisphere comparison method, while a
preferred direction ($b=-14.3^\circ \pm 10.1^\circ, l=307.1^\circ
\pm 16.2^\circ$) for the dipole fit method. This result indicates
that the dipole fit is more sensitive than the hemisphere comparison
method. 
\end{abstract}

\begin{keywords}
cosmology: theory - dark energy, Type Ia supernovae
\end{keywords}

\section{Introduction}
Einstein's general relativity and the cosmological principle are the
two key foundations in modern cosmology. Cosmologists
usually assumed that the general relativity is the perfect law of
gravity from small to large scales, which has been tested by many
tests in solar system and a few cosmological tests
(e.g.\citep{Zhang07}). The cosmological
principle~\citep{Weinberg08} assumes that the universe is
homogeneous and isotropic on a sufficiently large scale. In
practice, the homogeneity and isotropy are confirmed by a variety of
cosmological observations, such as cosmic microwave background
radiation (CMBR)~\citep{Hinshaw12}, the secondary effect of
CMB~\citep{Zhang11}, galaxy pairs~\citep{Mari12,Wang13} and the
large scale structure (LSS)~\citep{Sebastian11}. So far
there is no any conclusive evidence for an anisotropic cosmological
model.

However, a possible challenge to the cosmological principle
was reported in recent years. Schwarz \& Weinhorst (2007) claimed
that a statistically significant anisotropy of the Hubble diagram
was found at 2$\sigma$ level at $z<0.2$ by using SNe Ia data. SNe Ia
data has been examined previously to test the isotropy of the
universe~\citep{Kolatt01,Bonvin06,Gordon07,Schwarz07}. For
comparison, we can divide the previous studies into two approaches
as follows. (i) Local Universe Constraint is defined as searching
for preferred direction work with low redshift astronomical probes
(e.g.\citep{Colin11,Kalus12}). (ii) Non-local Universe
Constraint is defined as the study with intermediate and high
redshift data (e.g.\citep{Antoniou10,Cai12}), which
includes the redshift tomography analysis (dividing full sample into
different redshift bins).

Since there is no well accepted upper limit value of redshift about
how large the local universe is currently, we simply choose
$z_{local}\leq0.2$ in our classification. It is obvious that if a
preferred direction or any other kind of anisotropy really exists,
the physical origins may be different between local and non-local
universe. Various local effect can lead to anisotropy in local
universe, such as the bulk flow towards the Shapley supercluster
\citep{Colin11}. Thus, the explanation of local universe anisotropy
is complicate and subtle. But if the non-local universe anisotropy
was confirmed by observation, the standard cosmological model
($\Lambda$CDM) based on cosmological principle must be modified.  So
there are two merits of our classification. First, it provides the
difference of the probing scale. Second, it implicates the different
theoretical origins. At the same time, one can easily take the study
by model-independent manner in local universe constraint
\citep{Kalus12} and by model-dependent way in non-local universe
constraint \citep{Cai12}. In previous works, there are serious
differences and disagreements among the studies on the possible
cosmological anisotropy. Some works found no statistically
significant evidence for anisotropy using the SNe Ia data
\citep{Tomita01,Blomqvist08,Gupta08,Gupta10,Blomqvist10}. However,
many studies found that there is a statistically significant
anisotropy \citep{Schwarz07,Cooke10,Colin11} or a cosmological
preferred direction \citep{Antoniou10,Cai12,Mariano12,Cai13,Zhao13}.
A few works either gave no distinct results
\citep{Cooray10,Campanelli11} or argued that the anisotropic result
of local universe constraint is not contradiction to the
$\Lambda$CDM  model \citep{Kalus12}.

In this work, we search for a cosmological preferred
direction from the latest Union2.1 data for the first time. For the
anisotropic analysis, we adopt two typical and sophisticated
approaches which are hemisphere comparison \citep{Antoniou10} and
dipole fit \citep{Mariano12}. Since the preferred direction is
almost independent of isotropic dark energy models \citep{Cai12}, we
choose two simple cosmological models, $\Lambda$CDM and $w$CDM for
the hemisphere comparison approach, and $\Lambda$CDM for the dipole
fit. In the first approach, we use the matter density and the
equation of state of dark energy as the diagnostic qualities in the
$\Lambda$CDM and $w$CDM, respectively. In the second method, we
employ distance modulus as the diagnostic quality in $\Lambda$CDM
model.

The paper is organized as follows.We present the Union2.1
data and the two methods in section 2. Section 3 gives the numerical
results. We compare and discuss our results with other works in
section 4. Section 5 is a brief summary.

\section{THE DATA AND METHODS}

\subsection{The Union2.1 data and preliminary formulae}

SNe Ia are important probes of the evolution of the universe. In
this work, we use the Union2.1 sample which is a compilation
consisting of 580 SNe Ia. The redshift range is from 0.015 to 1.414
\citep{Suzuki12}. Comparing to the Union2 data, the updated Union2.1
data consists other 23 SNe Ia. Here we get the directions of Union2
data in the equatorial coordinates (right ascension and declination)
to each SN Ia from Blomqvist et al. (2010). We get the directions of
additional 23 datapoints from NED
website\footnote{http://ned.ipac.caltech.edu}.We also use the
Union2.1 table from the SCP
website\footnote{http://supernova.lbl.gov}, which includes
each SN Ia\rq{}s name, redshift, distance modulus and
uncertainties. We translated the equatorial coordinates of SNe Ia
to galactic coordinates $(l,b)$ in the galactic systems
\citep{Smith89}.

\begin{figure*}
\centering
\includegraphics[scale=0.44]{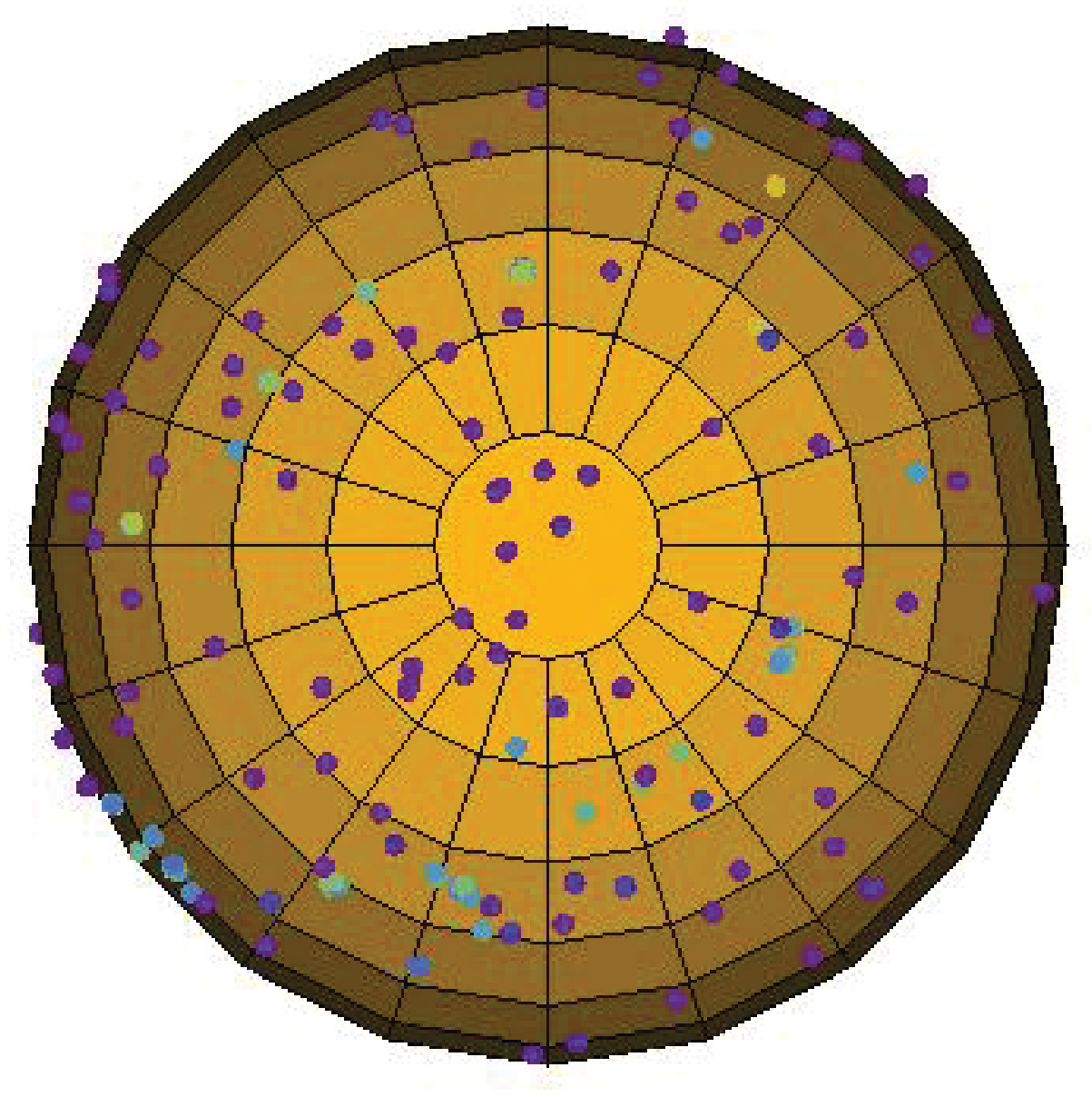}
\includegraphics[scale=0.55]{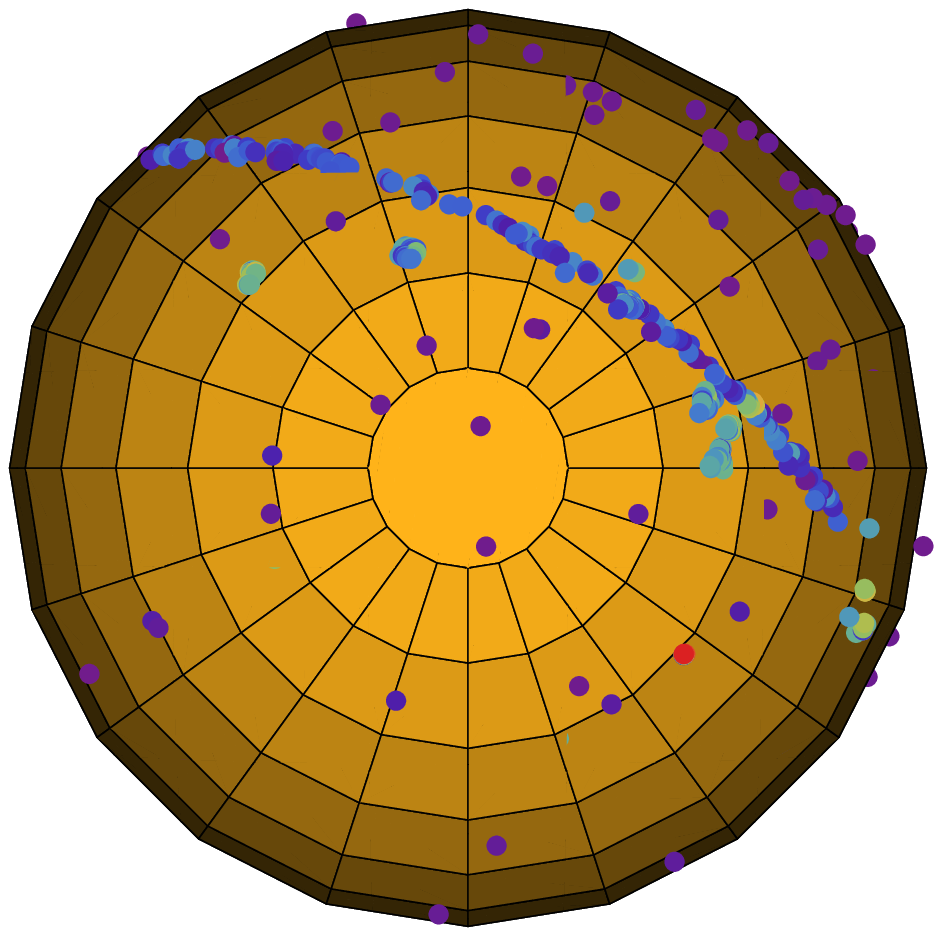}
\includegraphics[scale=0.7]{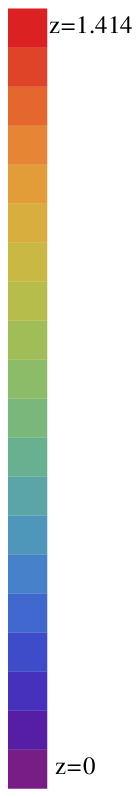}
\includegraphics[scale=0.44]{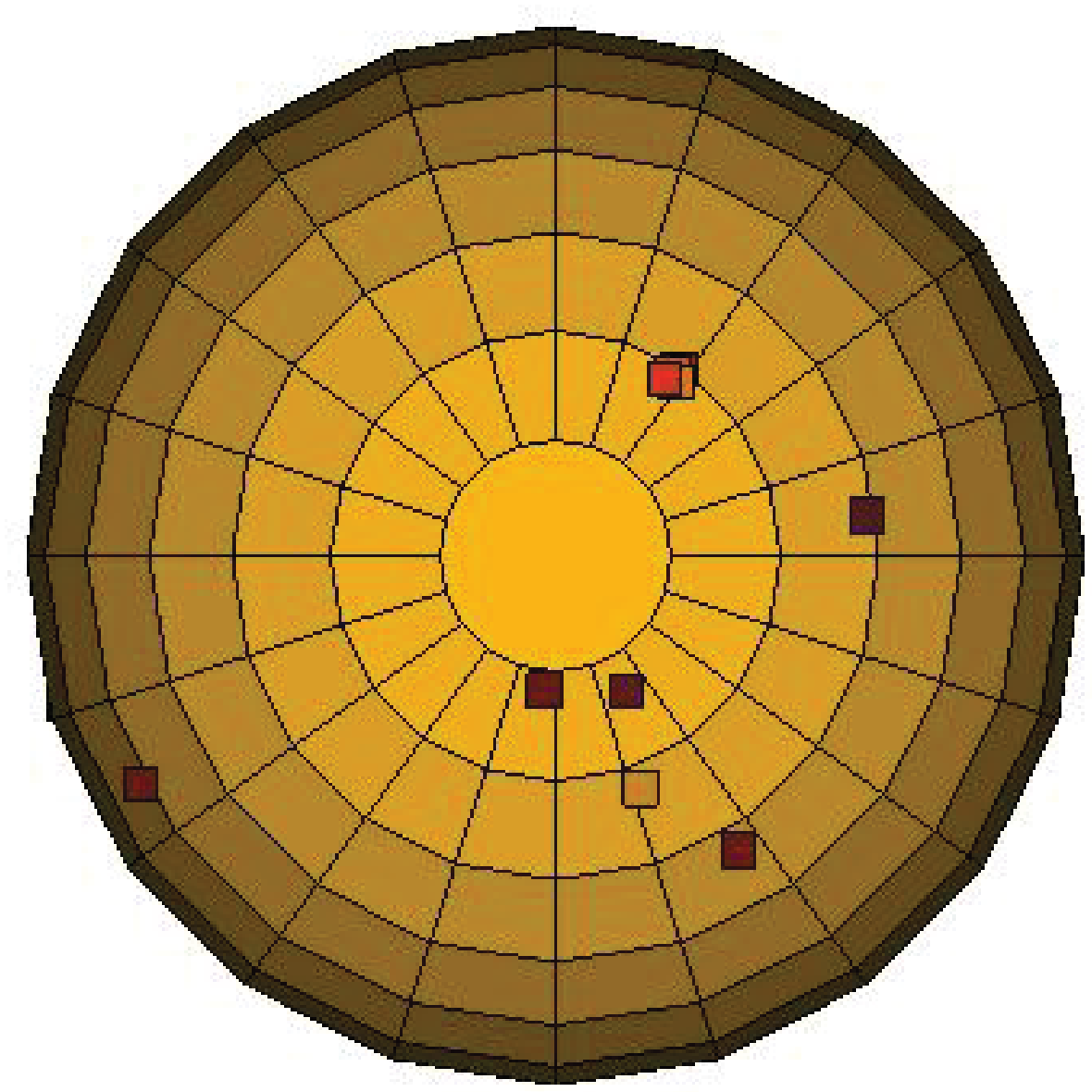}
\includegraphics[scale=0.55]{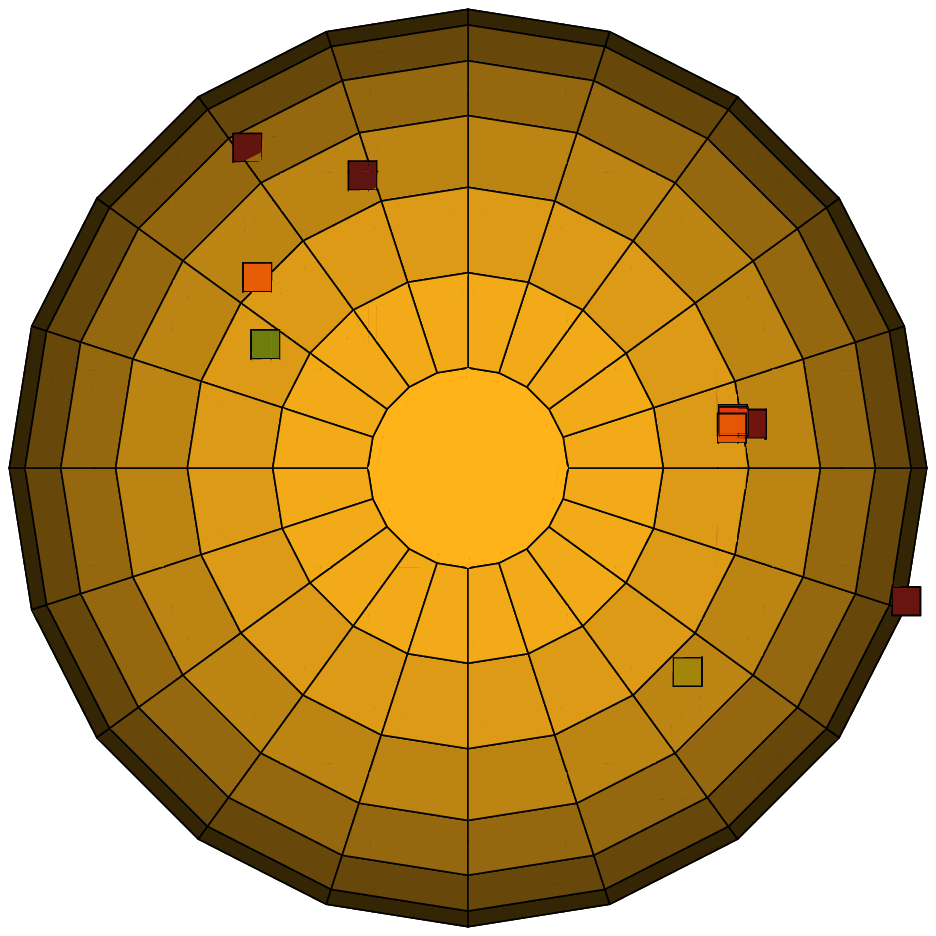}

\caption{(color online) The Union2 data (up panels) and
additional 23 data (bottom panels) in galactic coordinates. They are
shown with viewpoint above the north galactic equator and south
galactic equator in the left panel and right panel, respectively.
The color of each point corresponds to the redshift of each SN Ia.
}
 \label{fig:sn}
\end{figure*}

In Figure.\ref{fig:sn}, we show the angular distributions of
the Union 2.1 datapoints in galactic coordinates. The color
represents the value of redshift according to the legend on the
right. The figures are viewed above the north galactic equator and
south galactic equator in the left panel and right panel,
respectively. For avoiding confusion, we do\rq{}t show Union2 data
and additional 23 data on the same sphere. We show Union2 data in
top panels and additional 23 data in bottom panels, respectively.
Some of datapoints are nearly overlap in different redshift because
of the similar angular direction. It is obvious that the
distribution of additional 23 datapoints are slightly more isotropic
than the distribution of Union2 data.

We study the SNe Ia data in the classical way by applying the
maximum likelihood method. In a flat FLRW cosmological model, the
luminosity distance is
\begin{equation}
D_L(z)=(1+z)\int_{0}^{z}\frac{d{z}\rq{}}{E({z}\rq{})}.
\end{equation}
In the flat $\Lambda$CDM model, $E({z})$ can be parameterized by
\begin{equation}
E^2(z)=\Omega_{m0}(1+z)^3+(1-\Omega_{m0}),
\end{equation}
where $\Omega_{m0}$ is the matter density. For the $w$CDM model,
$E({z})$ is
\begin{equation}
E^2(z)=\Omega_{m0}(1+z)^3+(1-\Omega_{m0})(1+z)^{3+3w},
\end{equation}
where $w$ is the equation of state of dark energy.

We use the distance modulus of SN Ia data by minimizing the
$\chi^2$. The $\chi^{2}$ for SNe Ia is obtained by comparing
theoretical distance modulus
\begin{equation}
\mu_{th}(z)=5\log_{10}\big(D_L(z)\big)+\mu_{0},
\end{equation}
here,
\begin{equation}
\mu_{0}=42.38-5\log_{10}h
 \end{equation}
is a nuisance parameter. The theoretical model parameter
($\Omega_{m0}$ or $w$) is determined by minimizing the value of
$\chi^{2}$  with observed $\mu_{obs}$ of SNe Ia:
\begin{equation}
\chi_{\bf SN}^{2}(\Omega_{m0} ,\mu_{0})=\sum_{i=1}^{580}\frac{\Big(\mu_{obs}(z_{i})-\mu_{th}(\Omega_{m0} ,\mu_{0},z_{i})\Big)^{2}}{\sigma_{\mu}^{2}(z_i)}.
\end{equation}
Since the  nuisance parameter  $\mu_0$ is independent of the
dataset, we can expand $\chi_{\bf SN}^{2}$ with respect to $\mu_{0}$
\citep{Nesseris05}:
\begin{equation}
\chi_{\bf SN}^{2}=A-2\mu_{0}B+\mu_{0}^{2}C,\label{eq:expand}\end{equation}
 here \begin{eqnarray*}
A & = & \sum_{i=1}^{580}\frac{\big(\mu_{obs}(z_{i})-\mu_{th}(z_{i}, \mu_{0}=0)\big)^{2}}{\sigma_{\mu}^{2}(z_{i})},\\
B & = & \sum_{i=1}^{580}\frac{\mu_{obs}(z_{i})-\mu_{th}(z_{i}, \mu_{0}=0)}{\sigma_{\mu}^{2}(z_{i})},\\
C & = & \sum_{i=1}^{580}\frac{1}{\sigma_{\mu}^{2}(z_{i})}.\end{eqnarray*}
The value of Eq.~(\ref{eq:expand}) is minimum for $\mu_0=B/C$ at
\begin{equation}
\widetilde{\chi}_{\bf SN}^{2}=\chi_{\bf SN, min}^{2}=A-B^{2}/C,
\end{equation}
which is not rely on $\mu_{0}$.

\subsection{The hemisphere comparison approach}

Currently, it is not easy to find the angular dependence of
anisotropy at small scale with significant confidence level using
SNe Ia. The reason is that the number density of SNe Ia is
relatively low, particular in the tomography analysis. Thus, we
firstly employ the hemisphere comparison for searching the largest
possible anisotropy in the largest scale of $\pi/2$.  An early
similar research has been done to a CMB sky map analysis
\citep{Eriksen04}. The subsequent studies found one of the several
anomalies in the WMAP data (e.g.\citep{Copi07}). The
hemisphere comparison method was firstly proposed for searching
largest possible anisotropy with SNe Ia by Schwarz \& Weinhorst
(2007). It was further developed and used for finding the possibly
preferred direction \citep{Antoniou10,Cai12}.

In recent works, different cosmological parameters are chosen as the
diagnostic qualities, such as $\Omega_{m0}$ \citep{Antoniou10},
$q_0$ \citep{Cai12} and $H_0$ \citep{Kalus12}. Since the preferred
direction is weakly depended on dark energy models \citep{Cai12}, we
simply consider two cosmological models, such as $\Lambda$CDM and
$w$CDM models. We also adopt $\Omega_{m0}$ and $w$ for $\Lambda$CDM
and $w$CDM as the diagnostic qualities, respectively. It could be
convenient to compare previous results \citep{Antoniou10} with ours.

We review the procedure of hemisphere comparison method in
short \citep{Antoniou10}. (i) Generate a random direction with the
same probability in unit sphere.  (ii) Divide the dataset into two
subsets according to the sign of the product between the vector
generated in the step (i) and the unit vector describing the
direction of each SN Ia in the dataset. We can split the data in two
opposite hemispheres, denoted by up and down. (iii) Calculate the
best fit value of cosmological parameter on each hemisphere. (iv)
Repeat a large times from step (i) to step (iii), and search the
maximum normalized difference  for the full data, thus one can get
the preferred direction of maximum anisotropy.

One can get more details of this method from the two
references \citep{Antoniou10,Cai12}. Here, we just describe the
third step of this method,  which estimates the best parameter in
each hemisphere. The subscripts $u$ and $d$ represent the best
parameter fitting value in the `up' and `down' hemispheres,
respectively. For estimating $\Omega_{m0}$ in $\Lambda$CDM model, we
can define \citep{Antoniou10}
\begin{equation}
\label{eq1}
\delta=\frac{\Delta
\Omega_{m0}}{\bar\Omega_{m0}}=\frac{\Omega_{m0,u}-\Omega_{m0,d}}{(\Omega_{m0,u}+\Omega_{m0,d})/2}.
\end{equation}

For fitting $w$ in the $w$CDM model, we define the  relative
anisotropic level with the equation of state of dark energy as
\begin{equation}
\label{eq2} \delta\rq{}=\frac{\Delta w}{\bar
w}=\frac{w_u-w_d}{(w_u+w_d)/2},
\end{equation}
where $w_u$ and $w_d$ are the best fitting equation of state in the
`up' and `down' hemispheres, respectively. The number of random axes
should be more than the number of SNe Ia on each hemisphere.
For Union2.1 sample, the number of data points per
hemisphere is approximate 290, we choose 400 axes in this works. 
Since the hemisphere comparison approach is not pretty fine
and sensitive enough to particular types of anisotropy
\citep{Mariano12}, it is just a rough estimation for global
property. We only implement the non-local universe constraint
without redshift tomography if there is no any anisotropic signal in
global constraint with the full sample in all redshift ranges.

\subsection{The dipole fit approach}
Dipole anisotropic fit method  has been used for searching
the anisotropy of fine structure constant with quasars on
cosmological scale. Mariano \& Perivolaropoulos (2012) firstly
applied this method to anisotropic study using SNe Ia
\citep{Mariano12}. The main steps of the dipole fit method are shown
as follows:
\begin{itemize}
\item
Convert the equatorial coordinates of SNe Ia to galactic
coordinates.
\item
Give the Cartesian coordinates of unit vectors $\hat n_i$
corresponding each SN Ia with galactic coordinates $(l,b)$. So, we
obtain \be \hat n_i = \cos(b_i)  \cos(l_i) \hat i + \cos(b_i)
\sin(l_i) \hat j + \sin(b_i) \hat k  \label{hatni}. \ee
\item
Define the angular distribution model with dipole and monopole \be
(\frac{\Delta\mu}{\bar \mu})=d\cos\theta + m, \label{dipmod} \ee
where $\mu$ is distance modulus, $m$ and $d$ denote the monopole and
dipole magnitude, respectively, $\cos\theta$ is the angle with the
dipole axis defined by the vector \be
 \vec D \equiv c_1 \hat i + c_2 \hat j + c_3 \hat k \label{vecd}.
 \ee
So
 \be \hat n_i \cdot \vec D = d \cos \theta_i \label{acostheta}.
  \ee
Then, we can fit the SNe Ia data to a dipole anisotropy model
(\ref{dipmod}) using the maximum likelihood method by minimizing \be
\chi^2({\vec D},m)=\sum_{i=1}^{580}
\frac{\left[(\frac{\Delta\mu}{\bar \mu})_i - d \cos\theta_i  - m
\right]^2}{\sigma_i^2}. \label{chi2union} \ee
\item
At last, we can obtain the magnitude and direction of the best fit
dipole in galactic coordinates from the best fit $c_i$ coordinates
(e.g.\@ $d=\sqrt{c_1^2 +c_2^2+c_3^2}$). The corresponding $1\sigma$
errors are obtained using the covariance matrix approach.
\end{itemize}

\section{THE RESULTS}
\subsection{Results of hemisphere comparison method}

We apply the hemisphere comparison method using the latest
Union2.1 dataset. Generally, one can expect that we should get the
similar results with recent works from Union2 sample
\citep{Antoniou10,Cai12}. It is surprised that we get different
results compared with previous works. 

Table 1 shows our numerical results with the Union2.1 dataset, which
could be clearly compared with previous results shown in the second
row \citep{Antoniou10}. The 1 $\sigma$ error is propagated from the
uncertainties of the SNe Ia distance moduli. The superscript $Real$
and $Sim$ denote the maximum anisotropic values which are obtained
from real SNe Ia dataset and a typical isotropic simulated dataset,
respectively. The simulated isotropic dataset has been constructed
by replacing each real data distance modulus to a random number from
the normal distribution with mean and standard deviation obtained by
the best fitting value of $\mu_{th} (z_i)$ and by uncertainties of
the corresponding real data point, respectively. Comparing to the
result derived from Union2 dataset, it is clear that the maximum
anisotropy level is $0.31\pm0.05$ for the Union2.1 dataset. However,
the value is $0.35\pm0.05$ for simulation data, which is larger than
the one of real data. In this calculation, the same parameter and
cosmology model ($\Lambda$CDM) are used for the two different
datasets.

The maximum anisotropic value will convergence in calculations with
real data by enlarging the random selected axes, whereas it\rq{}s
precise value will be fluctuated in repeated estimations with
simulated data for random selected effect. In $w$CDM model
calculation, the value of Eq.(\ref{eq2}) is $0.27\pm0.07$ and
$0.37\pm0.07$ in real data and simulated data, respectively. The
value of real data is smaller than the one in $\Lambda$CDM fitting
($0.31\pm0.05$) for different cosmological parameter and model. The
value ($0.37\pm0.07$) in this simulation dataset is still larger
than the one in real data ($0.27\pm0.07$).

Although our results show that the maximum anisotropy level is lower
than simulation isotropic dataset from Union2.1 dataset, we still
process the same numerical experiments as shown in the Antoniou \&
Perivolaropoulos (2010).  The purpose is to answer whether the
maximum anisotropy level for real data is higher or lower than
statistical isotropy. This kind of numerical experiments is not
intend to identify the maximum anisotropic direction in standard 400
axes searching procedure. We only want to compare the real data with
the isotropic simulation data. It is important to repeat the
comparison many times (40 in our case) for acceptable statistics.
Because of the limitations of searching time, we adopt 10 axes for
employing fast-speed Monte Carlo experiments (Antoniou \&
Perivolaropoulos 2010). This numerical experiment is important
because of more fluctuated values with maximum anisotropy level in
the simulation data.

From a set of numerical experiments, we get different results from
Union2.1 dataset comparing with Union2 dataset in Table 2. The
$Real$ or $Sim$ denotes the number of cases which maximum
anisotropic value of real data is larger or smaller than that of
simulated data, respectively.  For Union2 dataset, there is about
$1/3$ of the numerical experiments with
$\delta_{max}^{Sim}>\delta_{max}^{Real} $, which means that the
anisotropy level was larger than the one of the isotropic simulation
data \citep{Antoniou10}. However, we find that the possibility of
real data and simulation data which have a larger maximum
anisotropic value is nearly equal. The results in Table 2 are not
consistent with the work of Antoniou \& Perivolaropoulos (2010). In
order to test the dependence on the number of axes, we increase the
random axes from 10 to 50. As shown in Table 2, our conclusion is
unchanged.

\begin{table}
\caption{The value of maximum anisotropy for Union2.1 dataset and
isotropic simulation dataset. The second row is the value calculated
from Union2 dataset \citep{Antoniou10}. } \label{tab:result0}
\begin{tabular}{cccccc}
\hline\hline
Model(Sample) & Diagnostic& $\delta_{max}^{Real}$ &$\delta_{max}^{Sim}$  \\\hline
$\Lambda$CDM(Union2)  &$\Omega_{m0} $& $0.43\pm0.06$ & $0.36\pm0.06$  \\\hline
$\Lambda$CDM(Union2.1)  &$\Omega_{m0} $& $0.31\pm0.05$  & $0.35\pm0.05$  \\\hline
$w$CDM (Union2.1) &$w$ & $0.27\pm0.07$ & $0.37\pm0.07$ \\\hline
\end{tabular}
\end{table}

\begin{table}
\caption{The results of 40 times numerical experiments for the value
of maximum anisotropy with Union2.1 dataset and isotropic simulation
datasets. The second row is the result from Union2 dataset
\citep{Antoniou10}. The Real or Sim denotes the number of cases
which maximum anisotropic value of real data is larger or smaller
than that of simulated data, respectively. } \label{tab:result1}
\begin{tabular}{cccccc}
\hline\hline
Model(Sample) & Axes$\times$All times& Real & Sim  \\\hline
$\Lambda$CDM(Union2)  &$10\times40$& 26 & 14  \\\hline
$\Lambda$CDM(Union2.1)  &$10\times40$ & 19 & 21 \\
 &$20\times40$ & 17 & 23\\
 &$30\times40$ & 20 & 20\\
  &$40\times40$ & 22 & 18\\
  &$50\times40$ & 18 & 22\\\hline
$w$CDM (Union2.1) & $10\times40$ & 18 & 22 \\
 &$20\times40$ & 19 & 21\\
 &$30\times40$ & 17 & 23\\
  &$40\times40$ & 21 & 19\\
  &$50\times40$ & 20 & 20\\\hline
\end{tabular}
\end{table}

Since there is no anisotropic signal in global constraint
with full Union2.1 data, we will not apply the redshift tomography
analysis in this work. We use tomography analysis in next subsection
which implements a more sensitive searching approach.

\subsection{Results of dipole fit method}

We study the latest Union2.1 dataset using the dipole fit
method, which includes non-local universe constraint and tomography
constraint. First, we report the result of non-local universe
constraint with full Union2.1 data. Then, we will show the local
universe constraint and tomography results. 

We find the direction of the dark energy dipole with full data
 \be
 b=-14.3^\circ \pm 10.1^\circ, l=307.1^\circ \pm 16.2^\circ.
 \ee
 The values of the dipole and monopole magnitudes are
\ba
d_{Union2.1}&=&(1.2\pm 0.5) \times 10^{-3}, \label{dedip}\\
m_{Union2.1}&=&(1.9\pm 2.1) \times 10^{-4} .\label{demon} \ea
The statistical significance of the dark energy dipole is
about at the $2\sigma$ level. The direction of Union2 dipole is
($b=-15.1^\circ \pm 11.5^\circ$, $l=309.4^\circ \pm
18.0^\circ$)\citep{Mariano12}, and the dipole and monopole
magnitudes are \ba
d_{Union2}&=&(1.3\pm 0.6) \times 10^{-3}, \label{dedip0}\\
m_{Union2}&=&(2.0\pm 2.2) \times 10^{-4}. \label{demon0} \ea The
statistical significance of the dark energy dipole is also at the
$2\sigma$ level using Union2, thus, our results are consistent with
Mariano \& Perivolaropoulos 2012. 

\begin{figure*}
\centering
\includegraphics[scale=1.35]{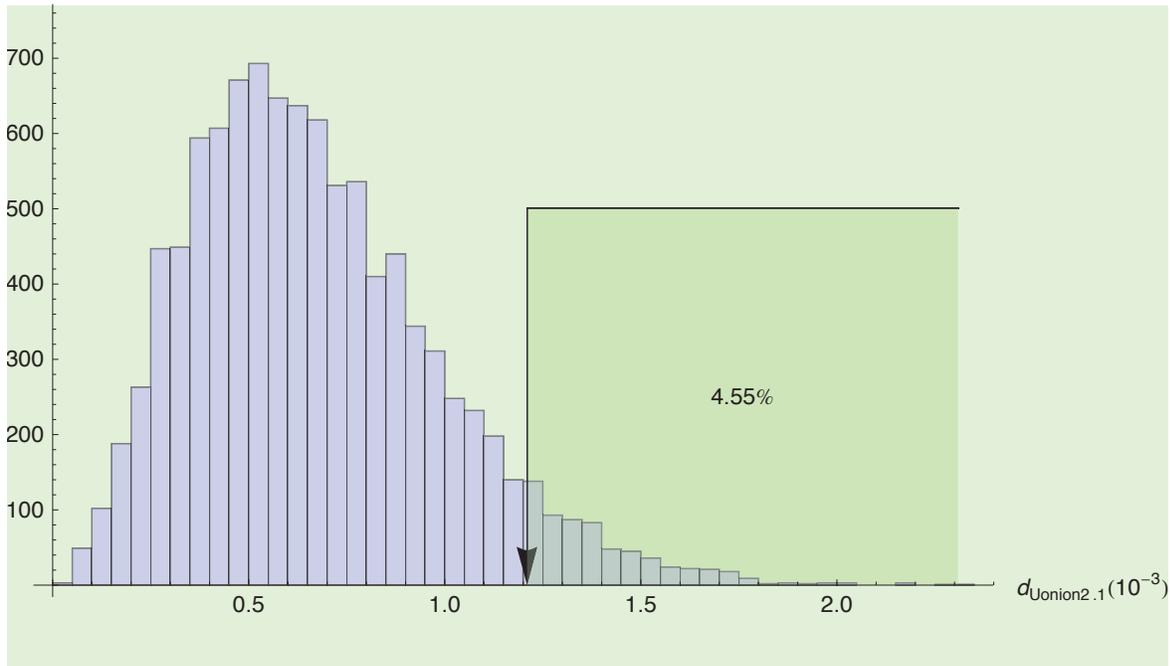}
\caption{(color online). Histogram of distribution
indicates the dark Energy dipole magnitudes from the Monte Carlo
simulation. The arrow position is the observed best fit value. The
deeper green region shows fraction of the Monte Carlo datasets that
give a dipole magnitude larger than the observed best fit one.}
\label{fig:MCdipole}
\end{figure*}

According to the dipole fit approach \citep{Mariano12}, we
determine the likelihood of the observed dark energy dipole
magnitude with performing a Monte Carlo simulation consisting of
$10^4$ Union2.1 datasets constructed under the assumption of
isotropic $\Lambda$CDM. The distance modulus of point $i$ is defined
as \be \mu_{MC}(z_i)=g(\bar \mu(z_i),\sigma_i), \label{mcsnia} \ee
where $g$ is the Gaussian random selection function
\citep{Mariano12}, and $\bar \mu(z_i)$ is the best fit distance
modulus of the Union2.1 full data in $\Lambda$CDM model at redshift
$z_i$. It is convenient to construct $\left(\frac{\Delta \mu
(z_i)}{\bar \mu (z_i)}\right)_{MC}$ for each Monte Carlo dataset and
search its best fit dipole direction and magnitude. In
Figure.~\ref{fig:MCdipole} we show the probability distribution of
the dark energy dipole magnitude along with the observed dipole
magnitude represented by an arrow. As expected from
Equation.~(\ref{dedip}) merely $4.55\%$ of the simulations had a
dark energy dipole magnitude bigger than the value in real dataset.
The result is consistent with Equation.~(\ref{dedip}) which
indicates that the statistical significance of the dark energy
dipole is about $2\sigma$. For the number of Monte Carlo simulation,
previous work proved that $10^4$ adopted as the number of simulated
datasets is enough to obtain a significant results
\citep{Mariano12}. 

We also take the redshift tomography analysis for indicating
these effects in different redshift ranges. We adopt two subsample
allocations similar as previous work based on Union2
\citep{Mariano12}, one is partitioning full sample with three
redshift bins and the other is changing the redshift upper limit. In
the first method, we divide full dataset into three redshift bins
which have nearly the same number of SNe Ia. Then we perform the
similar works as above in each bin and compare the results of each
bin with respect to the quality of data with errors, the dipole
magnitudes and the dipole directions. For the second method, we set
first and second subsample with an redshift upper limit consisting
of about a half of the full datapoints. Then we enlarge the redshift
upper limit properly so that the largest subsample almost includes
full dataset. We study each subsample of the six cumulative dataset
parts with the same procedure as above.

Table~\ref{tab:Union2.1Dipole} shows our results in
different redshift ranges with each subsample of Union2.1, which
includes our above non-local universe constraint in the second line.
It also shows the deviled method of redshift bins and the datapoints
number of each redshift bin. In 2nd to 5th columns, the results in
brackets are from Union2 data \citep{Mariano12}. In the last column,
the number in and out brackets is the datapoints of Union2 and
Union2.1, respectively. There is no additional datapoint from Union2
to Union2.1 in the redshift $0.14 < z \le 0.43$.  In each redshift
bin or range, we show the corresponding best fit monopole magnitude,
the dipole magnitude and the direction of the best fit dipole in
galactic coordinates. The uncertainties shown in
Table~\ref{tab:Union2.1Dipole} are calculated via the covariance
matrix approach. We have checked and confirmed that they are
consistent with the corresponding $1\sigma$ errors calculated from
the Monte Carlo simulations. All the results we reported here in the
Table~\ref{tab:Union2.1Dipole} are consistent with the results from
Union2 \citep{Mariano12}. We also find that the \lq\lq{}best\rq\rq{}
redshift bin with the smallest errors for the Union2.1 data is the
lowest redshift bin ($0.015 < z \le 0.14$), which is also similar to
previous work from Union2 \citep{Mariano12}.

\begin{table*}
\caption{The results of diploe fit approach including non-local,
local constraints and tomography analysis. In the 2nd to 5th
columns, we show the monopole magnitude, dipole magnitude and
direction from the estimation with Union2.1 (Union2) data in
different redshift ranges (1st column). Expect for the last column,
the results in brackets are the calculations from Union2 data
\citep{Mariano12} for comparison. In the last column, the number in
and out brackets represents the datapoints of Union2 and Union2.1,
respectively. There are the same datapoints between Union2.1 and
Union2 in the redshift $0.14 < z \le 0.43$. }\centering
\begin{tabular}{c|c|c|c|c|c}\hline\hline
 & $\frac{m_{U2.1}(m_{U2})}{10^{-4}}$ & $\frac{d_{U2.1}(d_{U2})}{10^{-3}}$ & $b_{d_{U2.1}}(b_{d_{U2}})$ & $l_{d_{U2.1}}(l_{d_{U2}})$ & U2.1(U2)\\
\hline
$0.015 \le z \le 1.414$  & $1.9\pm 2.1(2.0\pm 2.2)$ & 1.2 $\pm$ 0.5(1.3 $\pm$ 0.6) & $-14.3$ $\pm$ 10.1( $-15.1$ $\pm$ 11.5) & 307.1 $\pm$ 16.2(309.4 $\pm$ 18.0) & 580(577) \\
$0.015 < z \le 0.14$   & $2.5\pm 3.1$($2.6\pm 3.4$) & 1.5 $\pm$ 0.7(1.7 $\pm$ 0.8) & $-9.8$ $\pm$ 14.6($-10.1$ $\pm$ 15.1) & 304.3 $\pm$ 21.4(308.8 $\pm$ 22.8) &  193(184)\\
$0.14 < z \le 0.43$    & $2.6\pm 5.6$ & 1.2 $\pm$ 1.9 & $-10.7$ $\pm$ 28.7 & 291.4 $\pm$ 37.2 & 186(186) \\
$0.43 < z \le 1.414$     & $0.6\pm 3.7$($0.7\pm 4.3$) & 0.7 $\pm$ 0.7(0.9 $\pm$ 0.8) & $-25.9$ $\pm$ 29.7($-25.1$ $\pm$ 30.6) & 35.7 $\pm$ 73.1(34.3 $\pm$ 75.7) & 201(187) \\
$0.015 \le z \le 0.23$ & $3.2\pm 2.7$($3.3\pm 2.9$) & 1.6 $\pm$ 0.6(1.8 $\pm$ 0.7) & $-7.8$  $\pm$ 11.9($-8.5$  $\pm$ 12.4) & 300.3 $\pm$ 16.1(302.2 $\pm$ 16.6) & 248(239) \\
$0.015 \le z \le 0.31$ & $3.5\pm 2.7$($3.8\pm 2.9$) & 1.7 $\pm$ 0.6(1.9 $\pm$ 0.7) & $-6.8$  $\pm$ 11.1($-7.6$  $\pm$ 11.6) & 304.5 $\pm$ 13.6(307.0 $\pm$ 14.7) & 301(292) \\
$0.015 \le z \le 0.41$ & $2.8\pm 2.6$($3.0\pm 2.7$) & 1.6 $\pm$ 0.6(1.8 $\pm$ 0.7) & $-13.8$ $\pm$ 9.7($-14.4$ $\pm$ 10.3) & 301.5 $\pm$ 13.5(303.6 $\pm$ 14.4) & 361(352) \\
$0.015 \le z \le 0.51$ & $2.1\pm 2.5$($2.2\pm 2.6$) & 1.3 $\pm$ 0.6(1.4 $\pm$ 0.7) & $-14.1$ $\pm$ 12.1($-14.9$ $\pm$ 12.7) & 298.8 $\pm$ 17.8(301.3 $\pm$ 18.8) & 415(406) \\
$0.015 \le z \le 0.64$ & $2.0\pm 2.2$($2.1\pm 2.4$) & 1.3 $\pm$ 0.5(1.4 $\pm$ 0.6) & $-15.5$ $\pm$ 10.7($-16.0$ $\pm$ 11.0) & 302.4 $\pm$ 16.1(305.3 $\pm$ 16.9) & 474(464) \\
$0.015 \le z \le 0.89$ & $1.8\pm 2.1$($2.2\pm 2.3$) & 1.3 $\pm$ 0.5(1.4 $\pm$ 0.6) & $-14.8$ $\pm$ 10.0($-15.6$ $\pm$ 10.4) & 307.3 $\pm$ 15.2(309.8 $\pm$ 16.0) & 531(519) \\
\hline\hline
\end{tabular}
\label{tab:Union2.1Dipole}
\end{table*}

\section{DISCUSSION}
If a preferred direction or any other anisotropy could be
really confirmed at high significant level, particular in non-local
universe ($z>0.2$), we should abandon cosmological principle and
study the anisotropic cosmological models, e.g. vector field model,
Bianchi type I model or extended topological quintessence model
\citep{Mariano12}. A comprehensive introduction of various
observational probes on the preferred axis could be found in the
paper \citep{Perivolaropoulos11}. So far, the largest anisotropic
value ($>0.7$)  is given by Cai \& Tuo (2010)\rq{}s work from Union2
data, which adopted the deceleration parameter $q_0$ for estimation
via hemisphere comparison method. However, in all of previous works,
the significance of the violation to isotropic assumption of
cosmological principle are not high.  In fact, most of them are no
more than 2 $\sigma$ confidence level. Although people have proved
that the significance could be improved by correlations with other
preferred axes from different observations \citep{Antoniou10}, none
of them has been confirmed or has acceptable fundamental physical
theory. Since there are tensions in cosmological
constraints with different observations, maybe it need more works on
this issue with different probes.

There are merely adding 23 data points in this paper, thus
it is not reasonable that we get the different results compared with
previous works based on Union2. Interesting, we have the different
results by the hemisphere comparison method but obtain the same
results by the dipole fit method. There are three potential reasons
for such differences. The first is the possible tension between
Union2 and Union2.1.  Second, the different space distribution is
another factor. The third reason is the different method\rq{}s
sensitivity. For the data tension, recently, some other independent
works focused on constraining the dark energy model point out the
tension in datasets between Union2 and Union2.1
(e.g.\citep{Zhang12}). For the different distribution, we show that
the distribution of Union2.1 dataset is slightly better-distributed
than the one of Union2, this hint could be found in Figure
\ref{fig:sn}. However, Kalus et al.(2012) argued that the
non-uniform distribution has no significant impact on such
anisotropic estimation. Since their work is just local universe
constraint whereas our and the two other hemisphere comparison works
\citep{Antoniou10,Cai12} are non-local universe constraints, the
detailed analysis of the different anisotropic searching results by
hemisphere comparison method and other methods beyond the scope of
this work. The third aspect may be the main point, which
is caused by that the dipole fit method is more sensitive and
effective than hemisphere comparison method \citep{Mariano12}.
Generally,  our results confirm this idea with Union2.1 data. On the
other hand, although hemisphere comparison method is neither precise
nor perfect, it is really a model-independent approach. Since we
should define the angular distribution model as a fiducial model in
diploe fit method, it is much more model-dependent than hemisphere
comparison method. This situation is similar to the studies on dark
energy reconstruction. Many dark energy parameterizations could
enhance the precision of dark energy parameters constraint, but the
parameterizations also impose some bias on the exact evolution of
dynamical dark energy. Correspondingly, if we adopt any specific
angular distribution model in dipole fit method, such as the
Equation.\ref{dipmod} or the parameterization in Cai et al\rq{}s
work \citep{Cai13}, it may affect the result of the potential unbias
anisotropy of the universe. We will investigate this interesting
issue in future works. The high-redshift data, such as gamma-ray
bursts will be included~\citep{Basilakos08,Wang11,Wang011}.

\section{SUMMARY}

In this paper, we search for a preferred direction of
acceleration using the Union2.1 SNe Ia sample. At the beginning of
this paper we simply specify and classify previous searching works
into two types according to their sample\rq{}s redshift ranges. Many
authors found that a maximum (minimum) expansion (acceleration) in a
preferred direction by applying the hemisphere comparison method and
dipole fit method to SNe Ia sample. We use the latest Union2.1
sample on this study for the first time. We adopt two cosmological
models ($\Lambda$CDM, $w$CDM) for hemisphere comparison method and
$\Lambda$CDM model for dipole fit. In hemisphere comparison
approach, we use matter density and the equation of state of dark
energy as the diagnostic qualities in $\Lambda$CDM and $w$CDM
models, respectively. In dipole fit approach, we study the
fluctuation of distance modulus and take the tomography analysis
with different redshift ranges. Comparing with Union2, we find a
null signal for cosmological preferred direction by hemisphere
comparison method. But there is a preferred direction
($b=-14.3^\circ \pm 10.1^\circ, l=307.1^\circ \pm 16.2^\circ$) by
dipole fit approach. Our results confirm that the dipole fit method
is more sensitive than the hemisphere comparison method for the
searching of a cosmological preferred direction with SNe Ia.

\section*{ACKNOWLEDGMENTS}
We thank the referee Prof. Perivolaropoulos very much for the detail
and constructive suggestions which helped to improve the manuscript
significantly. We have benefited from reading the publicly available
codes of Antoniou \& Perivolaropoulos (2010) and Mariano \&
Perivolaropoulos (2012). Xiaofeng Yang gratefully acknowledges the
collaborating, long visiting and open fund provided by State Key
Laboratory of Theoretical Physics, Institute of Theoretical Physics,
Chinese Academy of Sciences, where the last revision of this
manuscript was completed in. This work is supported by the National
Basic Research Program of China (973 Program, grant 2014CB845800)
and the National Natural Science Foundation of China (grants
11373022, 11103007, and 11033002).


\begin{thebibliography}{36}


\bibitem[Antoniou \& Perivolaropoulos(2010)]{Antoniou10} Antoniou, A., Perivolaropoulos, L., \ 2010,  JCAP, 12, 012

\bibitem[Basilakos \& Perivolaropoulos(2008)]{Basilakos08} Basilakos S., Perivolaropoulos L., 2008, MNRAS, 391, 411

\bibitem[Blomqvist(2008)]{Blomqvist08}
Blomqvist. M., Mortsell. E.,  Nobili., S., \ 2008, JCAP, 06, 027

\bibitem[Blomqvist(2010)]{Blomqvist10}
Blomqvist. M., Enander. J., Mortsell. E., \ 2010, JCAP, 10, 018

\bibitem[Bonvin et al. (2006)]{Bonvin06} Bonvin, C., Durrer, R., Kunz, M., 2006, Phys. Rev. Lett.,
96, 191302

\bibitem[Cai \& Tuo(2012)]{Cai12}
Cai, R. G., Tuo, Z. L., \ 2012, JCAP, 02, 004

\bibitem[Cai et al.(2013)]{Cai13}
Cai, R. G., Ma, Y. Z.,  Tang, B., Tuo. Z. L., \ 2013, \prd, 87,
123522

\bibitem[Cooke \& Lynden(2010)]{Cooke10} Cooke, R., Lynden, B., D., \ 2010, \mnras, 401, 1409

\bibitem[Cooray et al.(2010)]{Cooray10} Cooray, A. R., Holz, D. E., Caldwell, R., \ 2010,  JCAP,  11, 015

\bibitem[Copi et al.(2007)]{Copi07} Copi, C. J., Huterer, D., Schwar, Z., Starkman, G. D.,  2007, \prd, 75,023507

\bibitem[Colin et al.(2011)]{Colin11} Colin, J., Mohayaee, R., Sarkar,S., Shafieloo, A.  2011, \mnras, 414,264

\bibitem[Campanelli et al.(2011)]{Campanelli11} Campanelli, L., Cea, P., Fogli, G. L., Marrone. A.,  2011, \prd, 83,103503

\bibitem[Eriksen et al.(2004)]{Eriksen04} Eriksen, H. K., Hansen, F. K., Banday. A. J., Gorski. K. M., Lilje. K. M.,  2004, \apj, 605, 14

\bibitem[Gordon et al. (2007)]{Gordon07} Gordon, C., Land, K. \& Slosar, A., 2007, Phys. Rev. Lett.,
99, 081301

\bibitem[Gupta et al.(2008)]{Gupta08} Gupta, S., Saini, T. D., Lskar,T., \ 2008, \mnras, 388,242

\bibitem[Gupta et al.(2010)]{Gupta10} Gupta, S., Saini, T. D. \ 2010, \mnras, 407,651

\bibitem[Hinshaw et al.(2013)]{Hinshaw12} Hinshaw, S. et al., \ 2013,
ApJS, 208, 19

\bibitem[Kalus et al.(2012)] {Kalus12} Kalus. B., Schwarz, D. J, Seikel, M., Wiegand, A. \ 2012, A\&A, 478, 719

\bibitem[Kolatt \& Lahav (2001)]{Kolatt01} Kolatt, T. S. \& Lahav, O., 2001, MNRAS, 323, 859.

\bibitem[Mariano \& Perivolaropoulos(2012)]{Mariano12}
Mariano, A., Perivolaropoulos, L., \ 2012, \prd, 86, 083517

\bibitem[Marinoni et al. (2012)]{Mari12} Marinoni, C., Bela, J., Buzzia, A., 2012, JCAP,
10, 036

\bibitem[Nesseris \& Perivolaropoulos(2005)]{Nesseris05} Nesseris, S., \&  Perivolaropoulos, L.,   \ 2005, \prd, 72,123519


\bibitem[Perivolaropoulos(2011)]{Perivolaropoulos11} Perivolaropoulos, L., \ 2011, arXiv:astro-ph/1104.0539


\bibitem[Schwarz \& Weinhorst(2007)] {Schwarz07} Schwarz, D. J., \& Weinhorst. B, \ 2007, A\&A, 474, 719


\bibitem[Sebastian et al.(2011)]{Sebastian11} Sebastian, T. G., Klypin. A., Primack. J., Romanowsky. R. J. \ 2011, \apj, 742, 1



\bibitem[Suzuki et al.(2012)]{Suzuki12} Suzuki, N.,et al.\ 2012, \apj, 746, 85

\bibitem[Smith(1989)]{Smith89}
Smith, D. P,  \ 1989,  Practicial astronomy with your calculator,
Cambridge University Press, Cambridge U.K.


\bibitem[Tomita(2001)]{Tomita01} Tomita, A., \ 2001, Progress in Theoretical Physics, 106, 929

\bibitem[Wang \& Dai (2011)]{Wang011} Wang F. Y., Dai Z. G., 2011,
A\&A, 536, A96

\bibitem[Wang \& Dai (2013)]{Wang13} Wang, F. Y., Dai, Z. G., 2013, MNRAS,
432, 3025.

\bibitem[Wang, Qi \& Dai (2011)]{Wang11} Wang F. Y., Qi S., Dai Z. G., 2011,
MNRAS, 415, 3423

\bibitem[Weinberg(2008)]{Weinberg08}
Weinberg, S., 2008, Cosmology, Oxford University Press, Oxford U.K.

\bibitem[Zhao et al.(2013)]{Zhao13} Zhao, W., Wu.P.X., Zhang.Y., \ 2013, IJMPD, 22, 1350060

\bibitem[Zhang et al.(2013)]{Zhang12} Zhang, J., Li, Y., Zhang, X., \
2013, EPJC, 73, 2280

\bibitem[Zhang et al.(2007)]{Zhang07} Zhang, P., Liguori, M., Bean, R., Dodelson, S.\ 2007, Phys. Rev. Lett., 99, 141302

\bibitem[Zhang \& Stebbins(2011)]{Zhang11} Zhang, P., Stebbins, A.\ 2011, Phys. Rev. Lett., 107,041301


\end{thebibliography}
\end{document}